\documentclass{mem}
\usepackage{natbib}\usepackage{txfonts}\usepackage{balance}
\usepackage{graphicx}
\usepackage[a4paper,breaklinks,dvipdfm]{hyperref}
\idline{00}{000}
\begin{document}
\def\teff{$T\rm_{eff }$}
\def\kms{$\mathrm {km s}^{-1}$}

\title{
Lithium evolution in the Milky Way discs: the view using large stellar samples
}

   \subtitle{}

\author{
R. \,Smiljanic\inst{1} 
          }

\institute{
Nicolaus Copernicus Astronomical Center, Polish Academy of Sciences, 
Bartycka 18, 00-716, Warsaw, Poland
\email{rsmiljanic@camk.edu.pl}
}

\authorrunning{Smiljanic}

\titlerunning{Lithium in the MW discs}

\abstract{
	This contribution presents an overview of the evolution of Li abundances in stars of the Galactic thin and thick discs, from the observational point of view. The focus is on Li abundances obtained by recent projects and surveys. To separate thin and thick disc stars, both chemical abundances, kinematics, and ages can be used. For thick disc stars, the Li evolution is uncertain, as differences appear depending on how the stars were separated. Nevertheless, it seems clear that most of the Galactic enrichment in Li takes place in the thin disc. Literature consensus also seems to exist regarding the decrease in the Li abundances of stars with metallicity above solar. A brief discussion is included on some of the uncertainties that should be kept in mind when trying to understand the Li observations. This review ends listing two interesting open questions regarding Li abundances in Milky Way disc stars.	
}
\maketitle{}

\section{Introduction}

The Milky Way  ``disc'' is itself divided into two apparently distinct structural components: the thin and the thick discs. The thick disc was first identified by \citet{1982PASJ...34..365Y} and \citet{1983MNRAS.202.1025G} in studies of the stellar vertical density distribution. The thick disc becomes the dominant population on heights above 1-2 kpc from the plane.

Thick disc stars have been found to be old ($>$ 10 Gyr) and to have enhanced [$\alpha$/Fe] ratio \citep[see, e.g.,][and references therein]{2017MNRAS.464.2610F}. Thin disc stars, instead, are younger and have lower levels of the [$\alpha$/Fe] ratio. Thin and thick disc stars also have different kinematic properties \citep[see, e.g.,][]{2005A&A...438..139S}. How the thick disc was formed is still under  discussion \citep[see, e.g.,][and references therein]{2014A&A...569A..13R}.

The distributions of ages, abundances of $\alpha$-elements, and kinematics can be used to tentatively separate thin disc stars from the thick disc ones. However, it is important to realize that while the distributions of those quantities are different in each disc population, some overlap can and does exist. These overlaps are one of the aspects that introduce uncertainties when assigning an individual star to a given disc component.

\section{Li in the Milky Way discs}

Ideally, to constrain the evolution of Li in the Milky Way, we would like to study the stellar Li abundances as a function of time. Stellar ages, however, are not easily derived with the required accuracy \citep[see][for a review]{2010ARA&A..48..581S}. Therefore, the evolution of Li abundances is mostly studied as a function of the stellar metallicity \citep[see][for an example]{2001ApJ...559..909T}.

\citet{2012ApJ...756...46R} seem to be the first to divide a sample of stars into thin and thick disc components for a comparative study of the Li abundances. The division was based on the kinematic properties of the stars computed with Hipparcos data and following the work of \citet{2003A&A...410..527B}.

In their results, looking at the upper envelope of the distribution of Li abundances, there is an increase in Li with increasing metallicity for thin disc stars. For thick disc stars, the trend of the upper envelope of Li abundances with [Fe/H] is flat. The abundance level is at similar values to the Spite plateau of halo stars \citep{1982A&A...115..357S}, suggesting that there was no evolution of Li in this stellar population. \citet{2012ApJ...756...46R} also investigated the trend of Li with stellar age, finding a clear continuity in the evolution of Li from the older thick disc stars to the younger thin disc ones. 

The upper envelope of the distribution of Li abundances is used in this type of analysis to account for possible Li depletion. This upper envelope is usually defined as the mean of the five or so stars with highest Li in a given range of metallicity. This is an attempt to take into account effects of stellar evolution on the surface abundance of Li. These effects are discussed in detail in many contributions of this conference (see, for example, those of J. Bouvier; S. Cassissi; N. Lagarde; S. Cristallo; B. Twarog; M. Carlos, among others). Following \citet{1988A&A...192..192R}, it has become usual to use the upper envelope of the distribution of Li abundances under the expectation that this is a good indicator of the original abundances of that element (but see discussion in Section 3).

\subsection{Li in thick disc stars}

For thin disc stars, the works that appeared later basically agreed that Li increases with increasing [Fe/H] (up to solar metallicity, see Section 2.2). For thick disc stars, however, the situation became unclear.

The next large samples of Li abundances where thin and thick disc stars were compared are those of \citet{2015A&A...576A..69D}, \citet{2016A&A...595A..18G}, \citet{2018A&A...610A..38F}, and \citet{2018A&A...615A.151B}. Their different findings highlight the difficulties in separating clear samples of thick disc stars. 

\citet{2015A&A...576A..69D} compared the chemical and kinematic approaches to separate the two stellar populations. In both cases, the authors found that the Li abundances in thick disc stars seem to decrease with increasing metallicity. This behavior would mean that there was no significant source of Li during the time of thick disc formation.

In \citet{2016A&A...595A..18G}, a study conducted within the AMBRE project \citep[an effort to analyse all stellar spectra of the ESO archive, see][]{2013Msngr.153...18D}, it was found instead that the Li abundances in thick disc stars increase with increasing [Fe/H]. The same result was found by \citet{2018A&A...610A..38F} using Li abundances derived by the \emph{Gaia}-ESO Survey \citep{2012Msngr.147...25G,2013Msngr.154...47R}. In these two investigations, thick disc stars were separated based on their chemical properties.

Afterwards, \citet{2018A&A...615A.151B} clearly demonstrated that the different ways used to define thick disc stars have a strong influence on how the trend looks like. These authors argue that the preferred way to identify thick disc stars should be based on their ages. In this case, when associating only stars older than 8 Gyr to the thick disc, they recover a trend of decreasing Li with increasing metallicity.

To settle the issue, it is important to confirm the results of \citet{2018A&A...615A.151B} using independent samples. This will be possible in the near future, when larger samples of Li abundances coming from surveys like GALAH (see contributions by Deepak and K. Lind) and \emph{Gaia}-ESO become available (see contributions by E. Franciosini; N. Sanna; and L. Magrini).

\subsection{Li abundances at metallicities above solar}

\citet{2015A&A...576A..69D} were the first to discuss a reverse in the trend of increasing Li with increasing [Fe/H]. In their sample, the maximum Li abundance is reached just before solar metallicity and it starts to decrease afterwards. \citet{2015A&A...576A..69D} tentatively explain this decrease as related to the deeper convective layers of more metal-rich stars.  

A similar reversal is present in the \citet{2012ApJ...756...46R} sample, although it was not discussed by those authors. The same trend has been confirmed in the samples analyzed by \citet{2016A&A...595A..18G}, \citet{2018A&A...610A..38F}, and \citet{2018A&A...615A.151B}, perhaps demonstrating that this is an ubiquitous feature.

The decrease of Li abundances in metal-rich stars has received considerable attention in the recent literature. Possible explanations include a scenario involving the migration of stars from the inner Galaxy \citep{2019A&A...623A..99G} or a lower rate of novae at high metallicities \citep{2019MNRAS.489.3539G}. Readers interested in this topic are referred to the contribution of G. Guiglion in this conference, to the two papers cited above, and to further discussion in \citet{2017A&A...606A.132P}, \citet{2019MNRAS.482.4372C}, and \citet{2019MNRAS.487.3946M}.

\begin{figure*}[t!]
	\centering
	\resizebox{0.75\hsize}{!}{\includegraphics[clip=true]{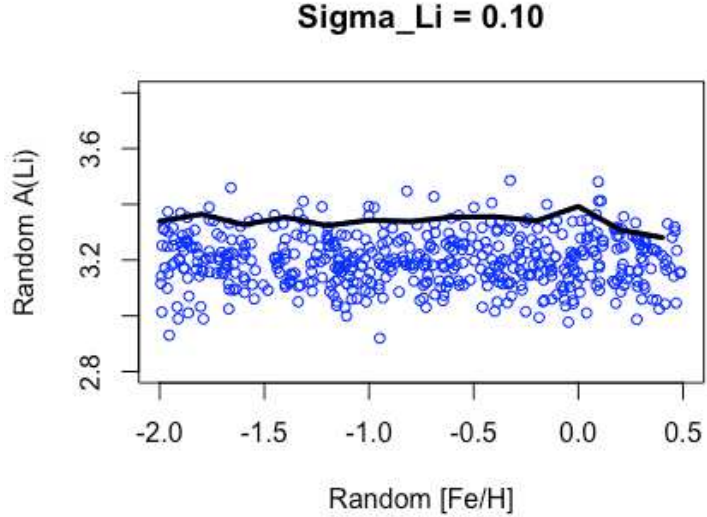}}
	\caption{\footnotesize
		Values representing a constant Li abundance of 3.2 dex affected by a random measurement error of $\pm$0.10 dex. The black line represents the upper envelope of the distribution, i.e., the mean of the five highest values in intervals of 0.2 dex in metallicity.
	}
	\label{fig:sigma}
\end{figure*}

\section{Uncertainties that affect the interpretation of the results}  

The discussion in the preceding Section covers all recent large stellar samples where thin and thick disc stars were separately discussed \citep[see, however,][for Li abundances in a sample of almost 2000 stars, but where such discussion was not made]{2017AJ....153...21L,2018AJ....155..111L}. In this Section, three sources of uncertainties, which can affect the interpretation of the evolution of Li abundances in the Galaxy, are mentioned.

\subsection{Random errors and outliers in automatic analyses}


When working with very large samples, a careful and detailed analysis of each single object in the sample is unfeasible. Thus, in most large spectroscopic surveys, automatic pipelines are used. In this context, quantifying both the systematic and random errors of the analysis becomes paramount for the proper use and understanding of the results that were obtained.

Quality control measures are of course part of most automatic analyses. However, to a certain extent, quality control is designed to catch the expected problems. Many of the unexpected issues might pass by unnoticed and make it to the final results. Perhaps that was not a dominant issue when the careful, ``by hand'', analysis of each object was possible. In results obtained by automatic pipelines, the pervasiveness of scatter and outliers is likely stronger.

The strategy of multiple analysis pipelines used by the \emph{Gaia}-ESO Survey \citep[see][]{2014A&A...570A.122S,2015A&A...576A..80L} has clearly demonstrated what was discussed above. The only way to understand the comparison of results from multiple pipelines is by taking into account the zero point offsets and the true random errors associated with each analysis. But detailed comparisons such as the ones done in \emph{Gaia}-ESO are not possible when a single pipeline is used. In such cases, quantifying the true magnitude of the systematic and random errors is more difficult.

In the context of this review, the discussion above is important because we base our understanding of the Li evolution on the behavior of the upper envelope of the abundance distribution. Usually, this upper envelope is computed taking into account a few (5-10) stars, those at the top of the distribution. The motivation is, of course, a physical one. The idea is to identify the stars that most likely retained their original Li abundance. But, by doing that in the era of large surveys, we do not take into account outliers and random errors.

Figure \ref{fig:sigma} is an attempt to illustrate the point. A number of random values were drawn from a Gaussian distribution of mean 3.2 and sigma 0.1, to represent Li abundances affected by a random error of $\pm$ 0.1 dex (which is a reasonable value to be expected from typical abundance analyses). These ``abundances'' were distributed in random metallicity values, drawn from an uniform distribution between $-$2.0 and +0.5. In the resulting plot, the five stars with higher Li, in intervals of 0.2 dex of [Fe/H], were chosen to compute the upper envelope. This upper envelope would suggest a level of Li between 3.3 and 3.4, in clear disagreement with the true value of 3.2. Real distributions of Li abundances might be affected just as in this example, causing a certain bias in our understanding of the Li evolution.

\subsection{Lithium depletion}

The surface depletion of Li is another important source of uncertainty that can bias the understanding of the Li evolution. In this case, the bias would go in the opposite direction with respect to the one discussed above.

In principle, there is no guarantee that the stars used to compute the upper envelope of the Li distribution have really retained their original surface Li abundance. In the \emph{Gaia}-ESO Survey, for example, it seems that the maximum Li detected in the sample is below the meteoritic value \citep[see][]{2018A&A...610A..38F}. 

To overcome this uncertainty, guidance from stellar evolution models is clearly needed. A variation in Li depletion is to be expected as a function of stellar metallicity, age and mass (at least). This variation should likely be taken into account for a proper quantification of the pace with which Li evolution proceeds. 

\subsection{Separating thin and thick disc stars}

As already discussed above, distinct ways to separate thin and thick disc stars can change the inferred behavior of the Li abundances. Interested readers are referred to \citet{2018A&A...615A.151B} for further details.

The additional point to be made is that not only the method used to classify the stars introduces an uncertainty. The errors of the quantities chosen to separate the stars can introduce extra uncertainties. 

This was also shown by \citet{2018A&A...615A.151B}. In their example, a change of 0.04 dex in the level of the [$\alpha$/Fe] ratio used to separate the stars, completely changed the inferred Li behavior. Similarly, it is to be expected that errors in the kinematics and ages can blur the attempts to distinguish between the populations.

\section{Open questions for future work}

To conclude the review, I mention here two challenging open questions that, to the best of my knowledge, have not been fully addressed from the observational point of view yet.

The first is whether there is a cosmic scatter of Li abundances? In other words, if stars with the same metallicity (or same age) can be formed with different initial Li abundances? As discussed before, at this moment, to determine the evolution of Li in the Milky Way, we are looking at the upper envelope of the distributions and attributing all the scatter below that to evolutionary effects that cause Li depletion.

However, just as for other elements, it is likely that there is a certain level of scatter of the Li abundance in the Galaxy. Understanding this scatter could give new important information to models of Galactic chemical evolution. Nevertheless, to detect and quantify a cosmic scatter, we would indeed need to infer what was the initial Li abundance of each star. And the obstacle to achieve that is of course the unknown Li depletion experienced by the star. 

On the other hand, it is clear that a lot of progress has been made towards understanding the multiple physical processes that change the stellar Li surface abundance. Eventually, it could become possible to reverse engineer the initial Li abundance of a star, if other relevant stellar properties like age, mass, and metallicity are well known. It might indeed already be feasible to get some qualitative idea about a possible Li scatter, at least on a statistical sense, if the proper tools are used.

The second open question concerns a possible Galactic radial gradient of Li abundances. Indeed, chemical evolution models seem to predict that a certain gradient should exist \citep{2012A&A...542A..67P}. The slope of this gradient could give information about the nucleosynthetic sources of Li and their dependence on star formation history and metallicity.

Measuring such gradient is of course very difficult. Lithium depletion is an important issue here again. In addition, there would be the need to observe large samples dwarfs, towards the Galactic center and anti-center. These are, however, fainter stars when compared to giants. It is thus more difficult to observe them at the large distances needed to build the radial gradient.

\section{Summary}

The large stellar samples, of several hundreds or thousands of stars, that are now becoming available thanks to spectroscopic surveys will be key to further advance the understanding of Li abundances in the Milky Way discs.

At this moment, there seems to be a consensus in the literature that the main Galactic enrichment of Li happens in the thin disc. Different observational studies also agree that there is a decrease in the level of Li abundances for stars more metal-rich than the Sun.

There is, however, doubt about the trend of Li abundances with [Fe/H] in stars of the thick disc. One important source of uncertainty is in how to define which stars belong to the thick disc in the first place. 

Quantifying the correct evolution of Li with [Fe/H] is difficult, as evolutionary Li depletion complicates the interpretation of the observations. Furthermore, in this era of automatic analysis pipelines, understanding the effect of random and systematic errors in the results is more important than ever.

Looking forward, among the questions concerning Li that remain to be explored are: i) is there a way to detect and quantify the existence of a cosmic Li scatter? and ii) what is the slope of the (current) Li radial gradient?
 
\begin{acknowledgements}
I acknowledge support from the National Science Center (NCN) of Poland through grant 2018/31/B/ST9/01469. I thank the organizers for the invitation to present this review and for the extremely interesting conference program.
\end{acknowledgements}

\bibliographystyle{aa}
\bibliography{../../../Biblio/smiljanic}

\end{document}